\documentclass[12pt,pra,showkeys]{revtex4}

\usepackage{amsmath}
\usepackage{graphicx}
\usepackage{psfrag}

\begin{document}

\title{Solutions of the Logarithmic Schr\"odinger Equation \\ in a Rotating Harmonic Trap}

\author{Iwo Bia{\l}ynicki-Birula}
\author{Tomasz Sowi{\'n}ski}
\affiliation{Center for Theoretical Physics, Polish Academy of Sciences\\
Lotnik{\'o}w 32/46, 02-668 Warszawa, Poland}

\begin{abstract}
We study the influence of the nonlinearity in the Schr\"odinger equation on the
motion of quantum particles in a harmonic trap. In order to obtain exact
analytic solutions, we have chosen the logarithmic nonlinearity. The unexpected
result of our study is the existence in the presence of nonlinearity of two or
even three coexisting Gaussian solutions.
\end{abstract}


\maketitle

\section{Introduction}
The nonlinear Schr\"odinger equation with the logarithmic nonlinearity (we use
the units $\hbar = 1$ and $m = 1$)
\begin{eqnarray}\label{lse}
 i\partial_t\psi({\bf r}, t) =  \left(-\frac{1}{2}
 \Delta + V({\bf r}, t) - b\log(\vert\psi({\bf r}, t)\vert^2/a^3)\right)
 \psi({\bf r}, t)
\end{eqnarray}
was introduced \cite{bbm1} long time ago to seek possible departures of quantum
mechanics from the linear regime. The parameter $b$ measures the strength of
the nonlinear interaction (positive $b$ means attraction) and $a$ is needed to
make the argument of the logarithm dimensionless --- it plays no significant
role since the change of $a$ results only in an additive constant to the
potential. In what follows, we shall absorb the parameter $a$ into the wave
function that amounts effectively to putting $a = 1$.

It has been proven in beautiful experiments with neutron beams
\cite{shimony,shull,gkz} that the nonlinear effects in quantum mechanics, if
they exist at all, are extremely small. The upper limit for the constant $b$
was determined to be $3.3\;10^{-15}$ eV. Thus, the applicability of the
logarithmic Schr\"odinger equation to the time evolution of wave functions
seems to have been ruled out. Nevertheless this equation, owing to its unique
mathematical properties, has been used in many branches of physics to model the
nonlinear behavior of various phenomena. It has been applied in the study of
dissipative systems \cite{hr}, in nuclear physics \cite{hefter}, in optics
\cite{keb,b4sc}, and even in geophysics \cite{dfgl}. In contrast to the
properties of  other nonlinear equations, the logarithmic Schr\"odinger
equation in any number of dimensions possesses analytic solutions, called
Gaussons in \cite{bbm2}. Gaussons represent localized nonspreading solutions of
the Gaussian shape. The internal structure of the Gaussons may also change in
time. The existence of these analytic solutions enables one to study in detail
the influence of nonlinearities. In this paper we focus our attention on the
behavior of the solutions of the logarithmic Schr\"odinger equation in a
rotating harmonic trap. The aim of our study was to see to what extent the
nonlinear interaction may change the dynamics and affect the stability of
solutions. Perhaps, our results will help to better understand the behavior of
the Bose-Einstein condensate in a rotating trap. Previous studies of these
problems (for example, \cite{rzs} and \cite{csbrd}) were often based on the
hydrodynamic equations and we plan in the future to express our results in
terms of the hydrodynamic variables.

\section{Formulation of the problem}
The logarithmic Schr\"odinger equation in a rotating trap has the form
\begin{equation}\label{nonleq}
 i\partial_t\psi({\mathbf r}, t)
 = \left(-\frac{1}{2}\Delta +
 \frac{1}{2}{\mathbf r}\!\cdot\!{\hat V}(t)\!\cdot\!{\mathbf r} -
 b\log(\vert\psi({\mathbf r}, t)\vert^2)\right)\psi({\mathbf r}, t),
\end{equation}
where the symmetric $3\times3$ matrix ${\hat V}(t)$ depends on time due to
rotation. In order to simplify the analysis of stability, we assume that the
trap is subjected to a uniform rotation and we shall use the coordinate system
co-rotating with the trap. In this manner the potential becomes
time-independent but due to rotation there appears an additional term in the
equation.
\begin{equation}\label{nonleq1}
 i\partial_t\psi({\mathbf r}, t)
 = \left(-\frac{1}{2}\Delta +
 \frac{1}{2}{\mathbf r}\!\cdot\!{\hat V}\!\cdot\!{\mathbf r}  -
 b\log(\vert\psi({\mathbf r}, t)\vert^2)
 - {\mathbf\Omega}\!\cdot\!{\bf M}\right)\psi({\mathbf r}, t),
\end{equation}
where ${\mathbf\Omega}$ is the vector of angular velocity and ${\bf M} = {\bf
r}\times{\bf p}$ is the operator of angular momentum. We shall seek the
solutions of Eq.~(\ref{nonleq1}) in the Gaussian form
\begin{eqnarray}\label{gausson}
 \psi({\mathbf r}, t)
 = N(t)e^{if(t)}\exp\left(-\frac{1}{2}
 {\tilde{\mathbf r}}\!\cdot\!({\hat A}(t)+i{\hat B}(t))\!\cdot\!
 {\tilde{\mathbf r}}(t) + i{\boldsymbol\pi}(t)\!\cdot\!{\bf r}\right),
\end{eqnarray}
where ${\tilde{\mathbf r}}={\mathbf r}-{\boldsymbol\xi}(t)$. The time-dependent
vectors ${\boldsymbol\xi}(t)$ and ${\boldsymbol\pi}(t)$ specify the position
and momentum of the center of mass of the Gaussian wave packet and the
time-dependent real symmetric matrices ${\hat A}(t)$ and ${\hat B}(t)$ specify
the shape and the internal motion of the wave packet, respectively. The two
real functions $N(t)$ and $f(t)$ define the normalization and the overall phase
of the Gausson. Substituting this Ansatz into Eq.~(\ref{nonleq1}), we arrive at
the following set of {\it ordinary} differential equations for all the
functions entering our formula (\ref{gausson})
\begin{eqnarray}\label{ode}
 \frac{d{\hat A}(t)}{dt} &=& {\hat B}(t){\hat A}(t) + {\hat A}(t){\hat B}(t)
 -\left[{\hat\Omega},{\hat A}(t)\right] ,\\
 \frac{d{\hat B}(t)}{dt} &=& {\hat B}(t)^2 - {\hat A}(t)^2 + {\hat V}
 + 2b{\hat A}(t) - \left[{\hat\Omega},{\hat B}(t)\right],\\
 \frac{d{\boldsymbol\xi}(t)}{dt} &=& {\boldsymbol\pi}(t) -
 {\boldsymbol\Omega}\times{\boldsymbol\xi}(t),\\
 \frac{d{\boldsymbol\pi}(t)}{dt} &=& - {\hat V}\!\cdot\!{\boldsymbol\xi}(t) -
 {\boldsymbol\Omega}\times{\boldsymbol\pi}(t),\\
 \frac{dN(t)}{dt} &=& \frac{1}{2}{\rm Tr}\{{\hat B}(t)\}N(t),\\
 \frac{df(t)}{dt} &=& -\frac{1}{2}\left({\rm Tr}\{{\hat A}(t)\}
 + {\boldsymbol\pi}(t)\!\cdot\!{\boldsymbol\pi}(t)
 - {\boldsymbol\xi}(t)\!\cdot\!{\hat V}\!\cdot\!{\boldsymbol\xi}(t)\right),
\end{eqnarray}
where the antisymmetric matrix ${\hat\Omega}$ and the components of the angular
velocity vector ${\boldsymbol\Omega}$ are related through the formula
$\Omega_{ij} = \epsilon_{ijk}\Omega^k$. Note, that the internal motion
(described by ${\hat A}(t)$ and ${\hat B}(t)$) completely decouples from the
motion of the center of mass (described by ${\boldsymbol\xi}(t)$ and
${\boldsymbol\pi}(t)$). In turn, the equations for the normalization factor and
the phase can be integrated after the internal and the center of mass motion
has been determined. This decoupling follows from the general theorem
\cite{gpv} and \cite{bb1} stating that from every solution of a nonlinear
Schr\"odinger equation in a harmonic potential (including time-dependent
potential) one may obtain a solution displaced by a classical trajectory fully
preserving the shape of the wave function.
\section{Solutions and their stability}

In what follows, for simplicity, we shall assume that the trap rotates along
one of its principal axis. In this case the motion in the direction
perpendicular to the rotation plane decouples and we are left with a
two-dimensional problem. In the stationary state of our system the center of
mass motion must be absent (${\boldsymbol\xi}(t)=0,{\boldsymbol\pi}(t)=0$) The
stationary state of the system is described by the wave function characterized
by the solution of the following two time-independent equations for two
$2\times 2$ matrices $A$ and $B$
\begin{eqnarray}
0 &=& {\hat B}{\hat A} + {\hat A}{\hat B} - \left[{\hat\Omega},{\hat
A}\right],\label{stat1}\\
 0 &=& {\hat B}^2 - {\hat A}^2 + {\hat V}
 + 2 b{\hat A} - \left[{\hat\Omega},{\hat B}\right].\label{stat2}
\end{eqnarray}
We shall seek the solutions of these equations in the coordinate frame in which
the matrix ${\hat V}$ is diagonal, ${\hat V} = {\rm
Diag}\{\omega_1^2,\omega_2^2\}$. We assume, for definitness, that $\omega_1 <
\omega_2$. It follows from Eqs.~(\ref{stat1}--\ref{stat2}) that in this frame
the matrix ${\hat A}$ is also diagonal and the matrix ${\hat B}$ is
off-diagonal. Finally, we are left with three equations for two matrix elements
$\alpha_1$, $\alpha_2$ of ${\hat A}$ and one matrix element $\beta$ of ${\hat
B}$
\begin{eqnarray}
 (\alpha_1+\alpha_2)\beta -(\alpha_1-\alpha_2)\Omega &=& 0,\label{eq1}\\
 \beta^2 - \alpha_1^2 + \omega_1^2 + 2 b\alpha_1 + 2\beta\Omega &=& 0,\label{eq2}\\
 \beta^2 - \alpha_2^2 + \omega_2^2 + 2 b\alpha_2 - 2\beta\Omega &=& 0.\label{eq3}
\end{eqnarray}
It follows from Eq.~(\ref{eq1}) that in the absence of rotation $\beta$ must
vanish and we obtain immediately two physically acceptable solutions of the
decoupled quadratic equations for the parameters $\alpha$
\begin{eqnarray}
 \alpha_1 &=& (\omega_1\sqrt{1 + b^2/\omega_1^2} + b),\label{sols01}\\
 \alpha_2 &=& (\omega_2\sqrt{1 + b^2/\omega_2^2} + b),\label{sols02}\\
 \beta &=& 0.\label{sols03}
\end{eqnarray}
The two remaining solutions yield negative values of the $\alpha$'s and must be
rejected. Thus, in the absence of rotation the nonlinearity modifies only the
size of the Gaussian wave function without introducing any significant changes.
Even for negative values of $b$ (nonlinear repulsion), stable solutions
described by (\ref{sols01}) and (\ref{sols02}) always exist, no matter how
strong is the repulsion.

Simple analytic formulas can also be obtained in the presence of rotation but
without nonlinearity. The formulas for the Gausson parameters read in this case
\begin{eqnarray}
 \alpha_1 &=& \frac{\sqrt{\omega_1^2+\omega_2^2+
 2\Omega^2 \pm 2\sqrt{(\omega_1^2-\Omega^2)(\omega_2^2-\Omega^2)}}}
 {1+\sqrt{(\omega_2^2-\Omega^2)/(\omega_1^2-\Omega^2)}},\label{sols11}\\
 \alpha_2 &=& \frac{\sqrt{\omega_1^2+\omega_2^2+
 2\Omega^2 \pm 2\sqrt{(\omega_1^2-\Omega^2)(\omega_2^2-\Omega^2)}}}
 {1+\sqrt{(\omega_1^2-\Omega^2)/(\omega_2^2-\Omega^2)}},\label{sols12}\\
 \beta &=& \Omega\frac{1-\sqrt{(\omega_2^2-\Omega^2)/(\omega_1^2-\Omega^2)}}
 {1+\sqrt{(\omega_2^2-\Omega^2)/(\omega_1^2-\Omega^2)}}.\label{sols13}
\end{eqnarray}
The values of $\alpha$ are real in the two regions of stability when $\Omega <
\omega_1$ (region 1) and $\Omega > \omega_2$ (region 2). The same regions of
stability were obtained in the analysis of the characteristic frequencies in
classical or quantum-mechanical center-of-mass motion \cite{bb1}. In the
formulas (\ref{sols11}) and (\ref{sols12}) the + and -- sign is to be chosen
for the region 1 and the region 2, respectively.

\begin{figure}[ht!]
\centering
\setlength{\abovecaptionskip}{-10pt}
\psfrag{lam\r}{$\alpha_1$,$\alpha_2$}
\psfrag{Om\r}{$\Omega$}
\psfrag{b\r}{\underline{$b=0$}}
\psfrag{o1\r}{$\Omega=\omega_1$}
\psfrag{o2\r}{$\Omega=\omega_2$}
\includegraphics{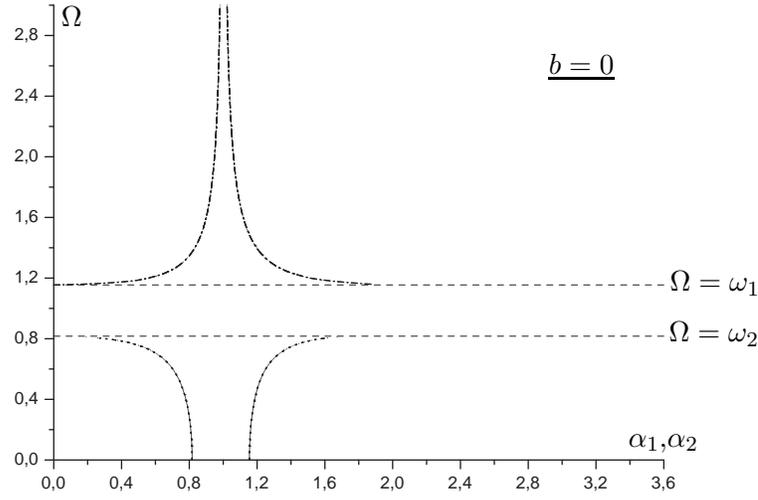}
\caption{This plot shows the values of $\alpha_1$ and $\alpha_2$ in the absence
of the nonlinearity. For each value of $\Omega$ in the stability regions there
is just one Gaussian wave function whose shape is described by the values of
$\alpha$'s.}
\label{fig:g1}
\end{figure}

\begin{figure}[ht!]
\centering
\setlength{\abovecaptionskip}{-10pt}
\psfrag{lam\r}{$\alpha_1$,$\alpha_2$}
\psfrag{Om\r}{$\Omega$}
\psfrag{b\r}{\underline{$b=1$}}
\psfrag{o1\r}{$\Omega=\omega_1$}
\psfrag{o2\r}{$\Omega=\omega_2$}
\includegraphics{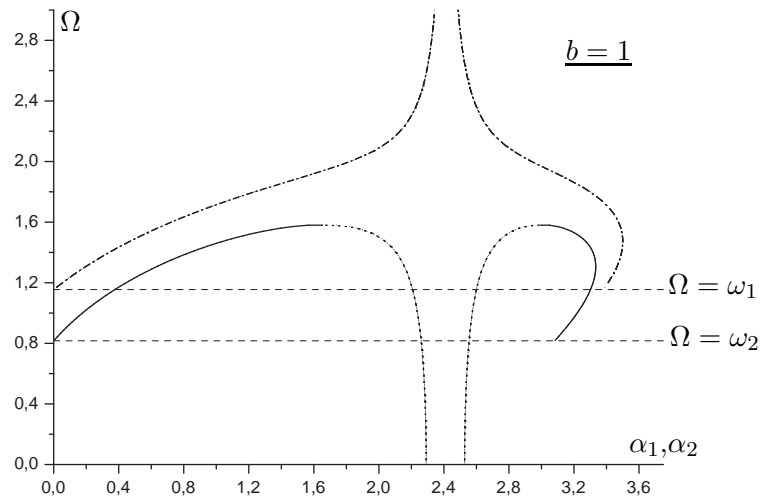}
\caption{This plot shows the values of $\alpha_1$ and $\alpha_2$ in the case of
attractive nonlinear interaction. For sufficiently large values of $b$, as in
this case, there are no regions of instability. For small and for large values
of $\Omega$ there is just one Gaussian wave function but for intermediate
values there are two or even three solutions. Matching pairs of $\alpha$'s are
distinguished by the lines of the same style: solid, dashed, and dotted.}
\label{fig:g2}
\end{figure}

\begin{figure}[ht!]
\centering
\setlength{\abovecaptionskip}{-10pt}
\psfrag{lam\r}{$\alpha_1$,$\alpha_2$}
\psfrag{Om\r}{$\Omega$}
\psfrag{b\r}{\underline{$b=-1$}}
\psfrag{o1\r}{$\Omega=\omega_1$}
\psfrag{o2\r}{$\Omega=\omega_2$}
\includegraphics{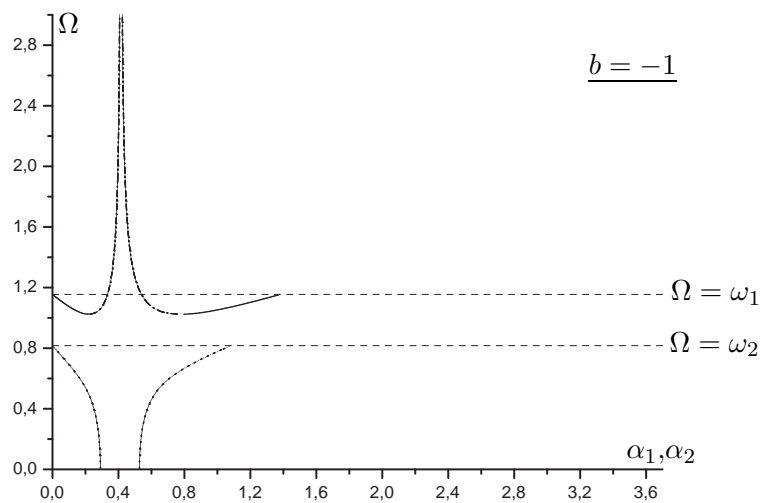}
\caption{This plot shows the values of $\alpha_1$ and $\alpha_2$ in the case of
repulsive nonlinear interaction. The upper region of stability is extended now
downwards as compared to the case without the nonlinear term. Moreover, there
are two solutions (solid and dashed lines) that coexist in the newly
established region of stability.} \label{fig:g3}
\end{figure}

In the presence of both rotation and nonlinearity the properties of solutions
change significantly. The most striking difference is the appearance of
additional stationary Gaussian solutions. This is an unexpected result because
in the linear theory a purely Gaussian shape always is found for {\em only one}
fundamental state of the system --- all other states have polynomial
prefactors. We have not been able to find closed expressions for the parameters
$\alpha$ and $\beta$, so we had to resort to numerical analysis of the
solutions of Eqs.~(\ref{eq1}--\ref{eq3}). We present our results in three plots
showing the calculated values of the parameters $\alpha_1$ and $\alpha_2$ that
determine the shape of the Gaussian wave function. These values are plotted as
functions of the angular velocity $\Omega$. In all plots we have fixed the trap
parameters to be $\omega_1 = \sqrt{2/3}, \omega_2 = \sqrt{4/3}$. We have chosen
three values of $b$ to describe the following characteristic cases. In
Fig.~\ref{fig:g1} we plot the values of $\alpha$'s without the nonlinear
interaction ($b = 0$). In Fig.~\ref{fig:g2} we added the attractive nonlinear
interaction ($b = 1$) and in Fig.~\ref{fig:g3} the repulsive nonlinear
interaction ($b = -1$).

\section{Conclusions}

Knowing the exact analytic form of the solutions of our nonlinear Schr\"odinger
equation we were able determine the influence of rotation and nonlinearity on
the stability of solutions. The unexpected result of our analysis is that the
repulsive interaction {\it expands} the region of stability. We have to admit,
however, that this may be true only for the special form of the nonlinearity:
the logarithmic nonlinearity.

\label{fig3}

\end{document}